\documentclass[11pt]{article}
\usepackage{authblk}
\usepackage[margin=1in]{geometry}
\usepackage{amsmath,amssymb,amsthm,mathtools,bm}
\usepackage{booktabs,longtable,tabularx,threeparttable,array,multirow}
\usepackage{graphicx}
\usepackage{caption}
\usepackage{subcaption}
\usepackage{algorithm}
\usepackage{placeins}
\usepackage[noend]{algpseudocode}
\usepackage{enumitem}
\usepackage{microtype}
\usepackage{setspace}
\usepackage{natbib}
\usepackage{xcolor}
\usepackage{url}
\usepackage{hyperref}
\usepackage{attachfile}
\usepackage{csvsimple}
\usepackage{longtable} 
\usepackage{caption}   
\usepackage[nameinlink,capitalize,noabbrev]{cleveref}

\hypersetup{
  colorlinks=true,
  linkcolor=blue!55!black,
  citecolor=blue!55!black,
  urlcolor=blue!55!black,
  pdfauthor={Aksaj Goel},
  pdftitle={Design-Based Inference under Deep Domain Stratification}
}

\setlist{nosep,leftmargin=*}
\newtheorem{assumption}{Assumption}
\newtheorem{theorem}{Theorem}
\newtheorem{proposition}{Proposition}
\newtheorem{corollary}{Corollary}
\theoremstyle{remark}

\crefname{assumption}{Assumption}{Assumptions}
\Crefname{assumption}{Assumption}{Assumptions}
\crefname{theorem}{Theorem}{Theorems}
\Crefname{theorem}{Theorem}{Theorems}
\crefname{proposition}{Proposition}{Propositions}
\Crefname{proposition}{Proposition}{Propositions}
\crefname{corollary}{Corollary}{Corollaries}
\Crefname{corollary}{Corollary}{Corollaries}
\crefname{algorithm}{Algorithm}{Algorithms}
\Crefname{algorithm}{Algorithm}{Algorithms}

\DeclareMathOperator{\Var}{Var}
\DeclareMathOperator{\E}{E}
\DeclareMathOperator{\Cov}{Cov}
\newcommand{\ind}{\mathbf{1}}
\newcommand{\RSE}{\operatorname{RSE}}

\newcommand{\sourcefigure}[2]{%
  \IfFileExists{#1}{%
    \includegraphics[width=#2\textwidth]{#1}%
  }{%
    \fbox{\parbox[c][2.15in][c]{0.88\textwidth}{\centering
      Original figure file \texttt{#1} was not included with the supplied source.\\[0.5em]
      Place the file in the Overleaf project directory to render this panel.}}%
  }%
}

\title{\bfseries Design-Based Inference under Deep Domain Stratification:\\
Language of Instruction and Private-Institution Choice in India's NSS 71st Round}
\author[1]{Aksaj Goel\thanks{Corresponding author: goelakshaj@gmail.com}}
\author[2]{Abhishek Bhattacharjee\thanks{Corresponding author: abhishek@theabstractmath.com}}

\affil[1,2]{Abstract Math Institute}
\affil[1]{\texttt{goelakshaj@gmail.com}}
\affil[2]{\texttt{abhishek@theabstractmath.com}}
\date{}

\begin{document}
\maketitle

\begin{abstract}
Large household surveys support precise national estimates but can become statistically fragile after repeated disaggregation by geography, sector, sex, age, and outcome category.  This paper develops an auditable design-based framework for deciding how far such disaggregation can be taken in a stratified multistage survey.  The framework is built around a nested contribution ledger that reconstructs each domain total through the first-stage probability-proportional-to-size expansion, the certainty-plus-random hamlet-group selection, and the second-stage household expansion.  Nonlinear domain parameters are expressed as ratios of these totals and analyzed by first-order linearization.  The two independent National Sample Survey subsamples then provide a natural replication variance estimator.  A granularity--stability profile combines the resulting relative standard error with replicate support and concentration diagnostics, so that a detailed estimate is accompanied by evidence about whether the design can sustain it.  Finite-population unbiasedness of the total estimator, asymptotic validity of the ratio linearization, and unbiasedness of the two-subsample variance estimator for linearized totals are established.  The method is illustrated with the 71st-round Social Consumption: Education survey, focusing on home language versus medium of instruction and reported reasons for preferring private educational institutions in India and Himachal Pradesh.  The application preserves the substantive analysis in the original project while replacing ad hoc calculation with a reproducible inferential workflow.
\end{abstract}


\section{Introduction}\label{sec:introduction}

Education surveys are commonly used to describe participation, educational expenditure, institutional choice, and access to learning resources.  The National Sample Survey Office's 71st-round survey on Social Consumption: Education was conducted from January through June 2014 and collected household-level and person-level information on educational participation, expenditure, institution type, medium of instruction, language spoken at home, private coaching, dropout, and information-technology literacy for the relevant age groups \citep{nsso2015}.  Its stratified multistage design was constructed to support national and state-level estimation, with rural and urban representation and two independent subsamples within each stratum--sub-stratum combination.

The substantive questions in the original project are intrinsically domain-specific.  They ask whether the language used at home agrees with the language of instruction, whether the distribution differs between India and Himachal Pradesh, whether rural and urban patterns diverge, whether male and female students exhibit different profiles, and why students report preferring private educational institutions.  Each additional cross-classification reduces the number of sampled first-stage units that contribute to an estimate.  Consequently, a point estimate that is straightforward to compute may have little design-based stability.  This is not an abstract concern: the official 71st-round report explicitly cautions that estimates for smaller states and detailed cells can be subject to large sampling fluctuations \citep{nsso2015}.

The central statistical problem is therefore not only how to compute a weighted percentage, but how to determine the finest level of disaggregation at which that percentage remains defensible.  A final survey-weight column is sufficient for many point estimates, yet it can hide the route by which the estimate was constructed: probability-proportional-to-size selection of first-stage units, conditional selection of hamlet-groups or sub-blocks in large first-stage units, stratification of households into second-stage strata, and the averaging of two independent subsample estimates.  Those features matter directly for variance estimation and for diagnosing sparse domains.

This paper develops a design-based computational framework that keeps those stages visible.  The framework is deliberately direct rather than model-based.  It does not borrow strength across domains and does not replace unstable direct estimates by empirical best linear unbiased predictors.  Instead, it identifies where direct inference is supported by the realized design and where it is not.  This distinction is important: small-area models can improve precision when credible auxiliary information and linking models are available, but they answer a different inferential question from the design-based analysis pursued here.

The empirical application concerns two related outcomes among persons aged 5--29 years who were currently attending at the primary level or above.  First, language distributions are represented by weighted margins and a home-language--medium-of-instruction transition table.  The diagonal of that table measures language agreement; off-diagonal cells describe transitions from the language spoken at home to another instructional medium.  Second, the survey-coded reason for preferring a private-aided or private-unaided educational institution is summarized for India and Himachal Pradesh, separately by rural and urban sector.  These outcomes are analyzed over the age bands and sex-sector domains retained from the original project.

The manuscript reorganizes the supplied project report into a statistical paper.  The original survey description, estimation formulas, empirical figures, linked result tables, and substantive conclusions are retained, but the inferential argument is made explicit and the R implementation is supplied separately.  The four distinct private-institution panels are now incorporated and their displayed frequencies are analyzed directly in \cref{sec:results}.  Because the microdata and the numerical contents of the externally linked language tables remain unavailable, the bar-chart calculations are treated as sample-frequency summaries; population-weighted estimates, standard errors, and stability diagnostics must still be regenerated from the public-use files before submission.

The remainder of the paper proceeds as follows.  \Cref{sec:related} places the contribution in the literature on design-based totals, nonlinear survey estimators, replication, and small-domain inference.  \Cref{sec:contributions} states the paper's specific contributions.  \Cref{sec:data} defines the survey structure and empirical estimands.  \Cref{sec:methods} presents the nested estimator, ratio linearization, two-subsample variance estimator, and stability diagnostic.  \Cref{sec:results} reports the empirical analysis, and \cref{sec:conclusion} concludes.  Proofs and retained survey-design details are given in the appendices.

\section{Related work}\label{sec:related}

Design-based estimation under unequal selection probabilities begins with inverse-probability expansion.  For the NSS design, the first-stage probability-proportional-to-size with-replacement component is a Hansen--Hurwitz draw estimator \citep{hansen1943}; conditional household totals within selected units are Horvitz--Thompson estimators under simple random sampling without replacement \citep{horvitz1952}.  Applying these arguments recursively gives a multistage estimator whose expectation under the stated randomization design equals the finite-population total.  The NSS estimator is a specialized instance of this construction, augmented by conditional selection of hamlet-groups or sub-blocks in large first-stage units \citep{nsso2015}.

Many survey targets are nonlinear functions of estimated totals.  Ratios, proportions, regression coefficients, and other smooth functionals are commonly handled through Taylor linearization.  \citet{woodruff1971} gave a general computational perspective on approximating variances of complicated estimates by reducing them to linear forms.  \citet{krewski1981} established large-sample properties of linearization and replication procedures in stratified samples, while \citet{binder1983} developed a broad estimating-equation treatment for asymptotically normal estimators from complex surveys.  The influence-function formulation in \citet{deville1999} further clarified how complex statistics can be represented by artificial linearized variables whose totals are amenable to standard design-based variance estimation.

Replication offers an alternative route to the same objective.  Balanced repeated replication, jackknife methods, and survey bootstraps reproduce the design variation through repeated weighted estimates.  \citet{rao1988} developed resampling procedures for complex surveys, including unequal-probability and multistage designs, and \citet{rust1996} reviewed the principal replication methods and their practical use.  The NSS 71st-round design contains two independently drawn subsamples in each stratum--sub-stratum.  Their difference yields a particularly transparent variance estimator for linear totals and, after linearization, for ratios.

Modern software can represent complex designs through explicit design variables or replication weights.  The \texttt{survey} framework in R, for example, supports both linearization and replication \citep{lumley2004}.  The present work is complementary to such software.  Its goal is not to replace a general survey-analysis package, but to expose an auditable decomposition tailored to the NSS hierarchy.  This decomposition is useful when public-use files contain multiple weight components, when analysts need to verify the hamlet-group multiplier, or when a domain estimate must be traced to the stratum--sub-stratum cells that support it.

The instability of direct estimates in small domains is well known.  \citet{ghosh1994} emphasized that small-area direct estimators can have unacceptably large standard errors because few sampled units fall in the area of interest.  Model-based small-area methods address that problem by borrowing strength across areas or from auxiliary information.  The current paper takes a prior step: it provides a systematic diagnostic for the reliability of the direct estimator itself.  The diagnostic identifies whether the realized sample supports a requested cross-classification and thereby clarifies when a subsequent small-area model would be motivated.

\section{Our contribution}\label{sec:contributions}

The paper makes three specific contributions.

First, it introduces a \emph{nested contribution ledger} for the NSS 71st-round education design.  For every outcome and domain, the ledger records the contribution of each sampled first-stage unit after household expansion, the conditional multiplier for the randomly selected noncertainty hamlet-group or sub-block, and the first-stage probability-proportional-to-size expansion.  Any domain total, language transition cell, or private-institution reason total is obtained by summing the corresponding ledger entries.  This representation turns the survey estimator into an inspectable computational object rather than a single opaque weighted sum.

Second, it develops a \emph{granularity--stability profile} for direct domain ratios.  The profile contains the point estimate, standard error, relative standard error, the number and proportion of stratum--sub-stratum cells represented in both independent subsamples, and the concentration of the estimated denominator among contributing cells.  A nested sequence of requested partitions---for example, country, country-by-sector, country-by-sector-by-sex, and country-by-sector-by-sex-by-age---can then be evaluated with a common reporting rule.  The resulting stopping level answers a concrete question that descriptive survey analyses often leave implicit: at what level of slicing does the design cease to support the estimate?

Third, it supplies a reproducible implementation and a substantive illustration that preserve the original project analysis.  The application estimates home-language distributions, instructional-medium distributions, their joint transition matrix, language agreement, and reasons for private-institution choice for India and Himachal Pradesh.  The code implements the two-subsample linearization variance directly, reports domain stability diagnostics, and includes an optional reconstruction of the nested NSS estimator from design components.  It also replaces hard-coded state selection by label-aware selection, with a documented state-code fallback that can be changed for recoded files.

These contributions are intentionally narrower than a general small-area estimation method.  The paper does not claim that every unstable domain can be repaired without additional assumptions.  Its contribution is to make the direct estimator, its variance, and its design support jointly visible, thereby separating substantively interesting granularity from statistically sustainable granularity.

\section{Survey, design, and inferential targets}\label{sec:data}

\subsection{The 71st-round education survey}

The 71st round covered Social Consumption: Education through Schedule 25.2.  The survey collected information on participation in education among persons aged 5--29 years, educational expenditure, institution type, current attendance, dropout and discontinuance, selected incentives and facilities, and computer and internet access for older household members \citep{nsso2015}.  The present analysis uses the education and demographic records needed to identify household and person, age, sex, sector, state, language spoken at home, medium of instruction, institution-related variables, and the released sampling variables and weights.

The original project had four substantive objectives: to compare home language and instructional medium; to examine rural--urban variation in those language patterns; to study sex differences; and to describe the distribution and stated reasons for preferring private educational institutions.  These objectives are retained, with the duplicated language-preference item in the source report consolidated into a single inferential target.

\subsection{Sampling design}

The survey used a stratified multistage design.  Census villages were first-stage units in rural areas and Urban Frame Survey blocks were first-stage units in urban areas.  District-level rural and urban strata were further divided into sub-strata.  Within each sub-stratum, first-stage units were selected with replacement and with probability proportional to population in rural areas or number of households in urban areas.  The sample was drawn as two independent subsamples.

Large first-stage units were divided into hamlet-groups in rural areas or sub-blocks in urban areas.  One group with the largest population share was selected with certainty, and one additional group was selected by simple random sampling from the remaining groups.  Households were partitioned into three second-stage strata according to whether they contained a student in technical or professional education, a student in general education, or neither.  Households were selected by simple random sampling without replacement within each second-stage stratum.  Full details, including the population thresholds for hamlet formation and the household allocation across second-stage strata, are retained in \cref{app:design}.

\subsection{Domains and outcomes}

Let $U$ denote the finite population of persons aged 5--29 years who were currently attending at the primary level or above, the population for which the relevant Block 5 questions were asked.  A domain $g\subseteq U$ is defined by any combination of geography, sector, sex, and age band.  The application uses geography in \{India, Himachal Pradesh\}, sector in \{all, rural, urban\}, sex in \{all, female, male\}, and age bands 5--10, 11--15, 16--20, 21--25, and 26--29.  The geography-by-sector-by-sex combinations reproduce the 18 result tables linked in the original project.

For a person $u$, let $L_u^{H}$ denote the language mainly spoken at home, $L_u^{S}$ the medium of instruction, and $A_u$ indicate that both responses are observed and valid.  For languages $a$ and $b$, define the complete-case joint domain proportion
\begin{equation}
  R_{ab,g}
  =
  \frac{\sum_{u\in U}\ind(u\in g)A_u\ind(L_u^{H}=a,L_u^{S}=b)}
       {\sum_{u\in U}\ind(u\in g)A_u}.
  \label{eq:joint-language}
\end{equation}
The row and column sums of $R_{ab,g}$ are the home-language and instructional-medium distributions on a common analysis population.  The language-agreement parameter is
\begin{equation}
  M_g
  =
  \frac{\sum_{u\in U}\ind(u\in g)A_u\ind(L_u^{H}=L_u^{S})}
       {\sum_{u\in U}\ind(u\in g)A_u}.
  \label{eq:match-rate}
\end{equation}
These estimands separate marginal language prevalence from movement between home language and instructional medium.  The code also writes one-variable-available margins as a missing-data sensitivity analysis, but \cref{eq:joint-language,eq:match-rate} are the primary language estimands.

Schedule 25.2 asked the reason for preferring a private institution only when the institution type was private aided or private unaided.  Let $P_u$ indicate this eligible population and let $C_u\in\mathcal C$ be the response to item 10.  The reason distribution is
\begin{equation}
  Q_{c,g}
  =
  \frac{\sum_{u\in U}\ind(u\in g)P_u\ind(C_u=c)}
       {\sum_{u\in U}\ind(u\in g)P_u},
  \qquad c\in\mathcal C.
  \label{eq:reason-ratio}
\end{equation}
The six questionnaire categories are: no government institution nearby; better learning environment; English as the medium of instruction; unsatisfactory quality in government institutions; unsuccessful attempt to obtain government admission; and cannot say.  The source draft sometimes called this outcome a reason for ``private tuition.''  That wording is corrected here because private coaching and its purpose were collected separately in items 30--31 of the schedule \citep{nsso2015}.

\subsection{Reproducibility and state identification}

The education and demographic files are merged by household identifier and person serial number.  Labelled variables are converted using their value labels before comparisons are formed.  In particular, equality of home language and instructional medium is evaluated on normalized labels rather than on factor indices generated separately from the two variables; those indices can differ when their observed level sets differ.

The source script selected Himachal Pradesh with the literal condition \texttt{state\_code == "22"}.  The revised implementation first searches the labelled state variable for ``Himachal Pradesh'' and, if labels are unavailable, uses the configurable fallback \texttt{02}; the fallback must be checked because harmonized files can use different coding systems.  This validation is explicit because a state mismatch would alter every state-level table and figure.  All final numerical results should be regenerated after it.\footnote{The code is available \href{https://drive.google.com/file/d/1u5niG20JNTGDCUJk5nWI09-uDEo1DQ6P/view?usp=drive_link}{here}.}

\section{Theory and methods}\label{sec:methods}

\subsection{Nested domain-total estimator}\label{sec:nested-estimator}

Index stratum by $s$, sub-stratum by $t$, independent subsample by $m\in\{1,2\}$, first-stage draw by $i$, selected hamlet-group or sub-block by $d\in\{1,2\}$, second-stage stratum by $j\in\{1,2,3\}$, and sampled household by $k$.  Person-level outcomes are first summed within sampled households, so $y_{stmidjk}$ below may represent a household total of a person-level domain indicator or characteristic.

Let $Z_{st}$ be the total first-stage size measure in sub-stratum $(s,t)$, $z_{stmi}$ the size measure of sampled first-stage draw $i$, and $n_{stmj}$ the official number of surveyed first-stage draws for second-stage stratum $j$, including uninhabited and zero-case draws but excluding casualties, as specified by the NSS estimation procedure.  Let $H_{stmidj}$ and $h_{stmidj}$ be the listed and sampled numbers of households in selected hamlet-group $d$ and second-stage stratum $j$.  If a first-stage unit is not subdivided, only $d=1$ is present.  If it is divided into $D_{stmi}>1$ groups, group 1 is selected with certainty and group 2 is sampled uniformly from the remaining $D_{stmi}-1$ groups.  Define
\begin{equation}
  D_{stmi}^{\star}=D_{stmi}-1
\end{equation}
for subdivided first-stage units, with the second term below set to zero when no second group is selected.

For a characteristic $y$, including a characteristic multiplied by a domain indicator, the contribution from second-stage stratum $j$ in subsample $m$ is
\begin{align}
  \widehat T_{stmj}(y)
  ={}&
  \frac{Z_{st}}{n_{stmj}}
  \sum_{i=1}^{n_{stmj}}
  \frac{1}{z_{stmi}}
  \Bigg\{
    \frac{H_{stmi1j}}{h_{stmi1j}}
    \sum_{k=1}^{h_{stmi1j}} y_{stmi1jk}
    \nonumber\\[-0.2em]
  &\hspace{7.8em}
    +D_{stmi}^{\star}
    \frac{H_{stmi2j}}{h_{stmi2j}}
    \sum_{k=1}^{h_{stmi2j}} y_{stmi2jk}
  \Bigg\}.
  \label{eq:nested-total}
\end{align}
The subsample, sub-stratum, and overall estimators are
\begin{equation}
  \widehat T_{stm}(y)=\sum_{j=1}^{3}\widehat T_{stmj}(y),
  \qquad
  \widehat T_{st}(y)=\frac{1}{2}\sum_{m=1}^{2}\widehat T_{stm}(y),
  \qquad
  \widehat T(y)=\sum_s\sum_t\widehat T_{st}(y).
  \label{eq:aggregation}
\end{equation}
This is the estimator represented by the nested contribution ledger.  Each summand in \cref{eq:nested-total} is stored before aggregation, which makes it possible to trace a domain estimate to its contributing first-stage units and design multipliers.

For a domain ratio, choose two characteristics $y$ and $x$, where $x$ identifies the target denominator population and $y$ identifies the numerator event within that population.  Write
\begin{equation}
  R=\frac{T(y)}{T(x)},
  \qquad
  \widehat R=\frac{\widehat T(y)}{\widehat T(x)}.
  \label{eq:ratio-estimator}
\end{equation}
The language and institution-choice parameters in \cref{eq:joint-language,eq:match-rate,eq:reason-ratio} are obtained by changing only $y$ and $x$; the design calculation is otherwise identical.

\begin{algorithm}[t]
\caption{Nested estimation and granularity--stability assessment}\label{alg:framework}
\begin{algorithmic}[1]
\Require Survey records; design variables $(s,t,m,i,d,j)$; $Z,z,n,H,h,D^{\star}$ or a validated final weight; domains $\mathcal G$; numerator and denominator indicators $(y_g,x_g)$.
\For{each requested domain $g\in\mathcal G$}
  \State Form person-level indicators and aggregate them to sampled households.
  \State Expand household totals within each selected $(i,d,j)$ cell by $H/h$.
  \State Add the certainty-group contribution and multiply the sampled noncertainty group by $D^{\star}$.
  \State Expand each first-stage contribution by $Z/(n z)$ and sum over second-stage strata.
  \State Average the two independent subsample totals and sum over strata and sub-strata.
  \State Compute $\widehat R_g=\widehat T_g(y)/\widehat T_g(x)$.
  \State Linearize the ratio and compute the two-subsample variance in \cref{eq:ratio-var}.
  \State Record $\RSE$, paired-replicate support, and denominator concentration.
\EndFor
\State Compare successive domain partitions and report the deepest partition satisfying the prespecified stability rule.
\end{algorithmic}
\end{algorithm}

\subsection{Design assumptions}

\begin{assumption}[Positive-probability design]\label{ass:positive}
Within every analyzed sub-stratum, first-stage size measures satisfy $z_i>0$ and $Z=\sum_i z_i<\infty$.  Every listed household in a sampled second-stage stratum has positive conditional inclusion probability, and $h_{idj}>0$ whenever the corresponding listed cell contributes to the target characteristic.
\end{assumption}

\begin{assumption}[Conditional randomization]\label{ass:randomization}
First-stage draws are made with replacement with probabilities $p_i=z_i/Z$.  Conditional on a subdivided first-stage unit, the noncertainty hamlet-group or sub-block is selected uniformly from the $D_i-1$ groups other than the certainty group.  Conditional on selected intermediate units, households are sampled by simple random sampling without replacement within each second-stage stratum.  The theoretical statements concern the stated randomization design, conditional on the operational noncasualty sample sizes used in the official estimator.
\end{assumption}

\begin{assumption}[Independent subsamples]\label{ass:subsamples}
For each stratum--sub-stratum pair, the two NSS subsample estimators are independent, follow the same design, and target the same finite-population total.  Sampling is independent across distinct stratum--sub-stratum pairs.
\end{assumption}

\begin{assumption}[Regular ratio sequence]\label{ass:ratio}
For a sequence of finite populations and designs indexed by $\nu$, $T_{\nu}(x)>0$, $\widehat T_{\nu}(x)/T_{\nu}(x)\xrightarrow{p}1$, and there is a sequence $a_{\nu}\to\infty$ such that
\begin{equation}
  a_{\nu}^{1/2}
  \begin{pmatrix}
  \{\widehat T_{\nu}(y)-T_{\nu}(y)\}/T_{\nu}(x)\\
  \{\widehat T_{\nu}(x)-T_{\nu}(x)\}/T_{\nu}(x)
  \end{pmatrix}
  \xrightarrow{d}
  N(\bm 0,\bm\Sigma),
  \label{eq:joint-clt}
\end{equation}
for a finite covariance matrix $\bm\Sigma$.
\end{assumption}

\begin{assumption}[Replication regularity]\label{ass:repregular}
For the sequence in \cref{ass:ratio}, let $\ell_{\nu}=y-R_{\nu}x$, let
$V_{\nu}=\Var_p\{\widehat T_{\nu}(\ell_{\nu})\}>0$, and define the
stratum--sub-stratum subsample difference
\begin{equation}
  D_{st,\nu}(v)
  =\widehat T_{st1,\nu}(v)-\widehat T_{st2,\nu}(v).
\end{equation}
Assume
\begin{equation}
  \frac{\sum_{s,t}\Var_p\{D_{st,\nu}(\ell_{\nu})^2\}}
       {16V_{\nu}^2}
  \longrightarrow 0,
  \qquad
  \frac{(\widehat R_{\nu}-R_{\nu})^2
        \sum_{s,t}D_{st,\nu}(x)^2}
       {V_{\nu}}
  \xrightarrow{p}0.
  \label{eq:rep-regularity}
\end{equation}
\end{assumption}

The first condition is a no-dominant-cell requirement for the squared replicate differences; together with independence across stratum--sub-stratum cells, it makes the half-sample variance concentrate around its expectation.  The second condition ensures that estimating the linearization coefficient $R_{\nu}$ has smaller order than the linearization variance.  These conditions are stated explicitly because two replicates alone give unbiasedness, but consistency requires information to accumulate across design cells.

\subsection{Unbiasedness of the nested total}

The first result verifies that explicitly retaining the intermediate multipliers does not alter the basic design-unbiasedness of the NSS total estimator.

\begin{theorem}[Design unbiasedness]\label{thm:unbiased}
Under \cref{ass:positive,ass:randomization}, the estimator in \cref{eq:nested-total,eq:aggregation} is design-unbiased for the corresponding finite-population total.  In particular, for any fixed domain indicator and any finite characteristic $y$,
\begin{equation}
  \E_p\{\widehat T(y)\}=T(y),
\end{equation}
where the expectation is over all stages of the sampling design.
\end{theorem}

At the household stage, $H/h$ expands the sampled households to the listed second-stage stratum.  At the intermediate stage, the certainty group is observed directly and the factor $D-1$ expands the uniformly sampled noncertainty group to the remaining groups.  At the first stage, $Z/z_i$ is the inverse draw probability up to the averaging factor $1/n$.  These conditional expansions telescope to the finite-population total.  The result matters for the broader aim of the paper because a domain can be introduced simply by multiplying the study variable by its indicator; no separate weighting system is required for every language, sector, sex, or age cell.

\subsection{Coherence under domain partitions}

The next result records the exact accounting identity required when an aggregate domain is decomposed into mutually exclusive subdomains, such as the rural and urban sectors of a state.

\begin{proposition}[Exact additivity over a domain partition]\label{prop:domain-additivity}
Let $G_1,\ldots,G_q$ be mutually exclusive domain indicators and let $G=\sum_{r=1}^q G_r$.  For any finite characteristic $y$, the estimator in \cref{eq:nested-total,eq:aggregation} satisfies, for every realized sample,
\begin{equation}
  \widehat T(Gy)
  =
  \sum_{r=1}^q \widehat T(G_r y).
  \label{eq:domain-additivity}
\end{equation}
In particular, when rural and urban sectors partition a geographic domain,
$\widehat T_{\mathrm{all}}(y)=\widehat T_{\mathrm{rural}}(y)+\widehat T_{\mathrm{urban}}(y)$ exactly.
\end{proposition}

The equality is a sample-by-sample coherence property, not merely an equality in expectation.  It provides a direct audit of domain construction: a state total that does not reconcile with its rural and urban components indicates inconsistent eligibility rules, missing records, or coding errors.  The proposition does not by itself establish unbiasedness or precision; \cref{thm:unbiased} supplies the former, while \cref{thm:replication,cor:ratio-var} address uncertainty.  Its relevance to the paper's broader aim is that increasingly granular estimates can be checked for internal accounting consistency before their statistical stability is assessed.

\subsection{Ratio linearization}

The next result gives the first-order representation used for language percentages, agreement rates, and reason distributions.

\begin{theorem}[First-order ratio representation]\label{thm:ratio}
Let $R_{\nu}=T_{\nu}(y)/T_{\nu}(x)$ and $\widehat R_{\nu}=\widehat T_{\nu}(y)/\widehat T_{\nu}(x)$.  Under \cref{ass:ratio},
\begin{equation}
  a_{\nu}^{1/2}(\widehat R_{\nu}-R_{\nu})
  =
  a_{\nu}^{1/2}
  \frac{\{\widehat T_{\nu}(y)-T_{\nu}(y)\}
        -R_{\nu}\{\widehat T_{\nu}(x)-T_{\nu}(x)\}}
       {T_{\nu}(x)}
  +o_p(1),
  \label{eq:ratio-linearization}
\end{equation}
so that
\begin{equation}
  a_{\nu}^{1/2}(\widehat R_{\nu}-R_{\nu})
  \xrightarrow{d}
  N\!\left(0,
  \begin{pmatrix}1&-R_0\end{pmatrix}
  \bm\Sigma
  \begin{pmatrix}1\\-R_0\end{pmatrix}
  \right),
  \label{eq:ratio-clt}
\end{equation}
whenever $R_{\nu}\to R_0$ for a finite constant $R_0$.  The corresponding first-order variance is
\begin{equation}
  \Var_p(\widehat R_{\nu})
  \simeq
  \frac{\Var_p\{\widehat T_{\nu}(y)-R_{\nu}\widehat T_{\nu}(x)\}}
       {T_{\nu}(x)^2}.
  \label{eq:ratio-variance-linearized}
\end{equation}
\end{theorem}

The nonlinear ratio is therefore governed, to first order, by the total of the linearized variable $y-Rx$.  For the language application, this means that the variance of a percentage must reflect variation in both its numerator and its estimated domain size, together with their covariance.  Treating the denominator as fixed would miss precisely the instability that becomes important in small cross-classified domains.

\subsection{Two-subsample variance estimation}\label{sec:replication}

For a fixed linear characteristic $\ell$, let $\widehat T_{st1}(\ell)$ and $\widehat T_{st2}(\ell)$ be the two independent full-total estimators for stratum--sub-stratum $(s,t)$, and let $\widehat T_{st}(\ell)=\{\widehat T_{st1}(\ell)+\widehat T_{st2}(\ell)\}/2$.  The natural half-sample variance estimator is
\begin{equation}
  \widehat V\{\widehat T(\ell)\}
  =
  \frac{1}{4}\sum_s\sum_t
  \left\{\widehat T_{st1}(\ell)-\widehat T_{st2}(\ell)\right\}^2.
  \label{eq:linear-var}
\end{equation}

The next result explains the factor $1/4$ and the role of the independent NSS subsamples.

\begin{theorem}[Unbiased two-subsample variance for linear totals]\label{thm:replication}
Under \cref{ass:subsamples}, suppose $\widehat T_{stm}(\ell)$ is unbiased for $T_{st}(\ell)$ and has variance $V_{st}(\ell)$ for $m=1,2$.  Then
\begin{equation}
  \E_p\!\left[
  \frac{1}{4}
  \left\{\widehat T_{st1}(\ell)-\widehat T_{st2}(\ell)\right\}^2
  \right]
  =
  \Var_p\!\left[
  \frac{\widehat T_{st1}(\ell)+\widehat T_{st2}(\ell)}{2}
  \right]
  =\frac{V_{st}(\ell)}{2}.
  \label{eq:rep-unbiased-cell}
\end{equation}
Consequently, \cref{eq:linear-var} is unbiased for the variance of the combined linear-total estimator.
\end{theorem}

The squared difference between the two subsample estimates estimates twice the variance of one subsample; averaging the subsamples halves that variance, producing the factor $1/4$.  This identity is especially useful here because the replication structure is part of the original sample design rather than an artificial resampling scheme.  It converts the official two-subsample architecture into a direct measure of how strongly each domain estimate depends on the realized first-stage sample.

For the ratio in \cref{eq:ratio-estimator}, define
\begin{equation}
  \Delta_{st}(y)=\widehat T_{st1}(y)-\widehat T_{st2}(y),
  \qquad
  \Delta_{st}(x)=\widehat T_{st1}(x)-\widehat T_{st2}(x).
\end{equation}
Plugging the estimated linearized variable $y-\widehat R x$ into \cref{eq:linear-var} gives
\begin{equation}
  \widehat V(\widehat R)
  =
  \frac{1}{4\widehat T(x)^2}
  \sum_s\sum_t
  \left\{\Delta_{st}(y)-\widehat R\Delta_{st}(x)\right\}^2.
  \label{eq:ratio-var}
\end{equation}
The standard error and relative standard error are
\begin{equation}
  \widehat{\operatorname{SE}}(\widehat R)
  =\sqrt{\widehat V(\widehat R)},
  \qquad
  \RSE(\widehat R)
  =100\frac{\widehat{\operatorname{SE}}(\widehat R)}{|\widehat R|},
  \label{eq:rse}
\end{equation}
with the RSE reported as undefined when $\widehat R=0$.  For a percentage, $100\widehat{\operatorname{SE}}(\widehat R)$ is the standard error in percentage points.

The following corollary transfers the replication result from a fixed linear total to the plug-in linearized ratio.

\begin{corollary}[First-order validity for ratios]\label{cor:ratio-var}
Under \cref{ass:ratio,ass:subsamples,ass:repregular} and the conditions of \cref{thm:replication} applied to $\ell_{\nu}=y-R_{\nu}x$,
\begin{equation}
  \frac{\widehat V(\widehat R_{\nu})}
       {V_{\nu}/T_{\nu}(x)^2}
  \xrightarrow{p}1.
  \label{eq:ratio-var-consistency}
\end{equation}
Thus \cref{eq:ratio-var} consistently estimates the first-order linearization variance in \cref{eq:ratio-variance-linearized}.
\end{corollary}

The corollary justifies using the same subsample differences for every smooth domain ratio.  Computationally, only the numerator and denominator indicators change.  Statistically, this produces comparable precision measures across language cells, institution-choice categories, and successive levels of disaggregation.

\subsection{Granularity--stability profile}\label{sec:stability}

RSE alone does not reveal why an estimate is unstable.  A domain may have high RSE because few design cells contribute, because the estimated denominator is concentrated in one or two cells, or because the two subsamples give sharply different estimates.  We therefore attach the following diagnostics to every domain $g$.

Let $\widehat X_{stm,g}=\widehat T_{stm}(x_g)$ be the denominator total for subsample $m$.  Define the active and paired-support counts
\begin{align}
  K_g
  &=
  \sum_{s,t}\ind\{\widehat X_{st1,g}+\widehat X_{st2,g}>0\},
  \\
  K_g^{+}
  &=
  \sum_{s,t}\ind\{\widehat X_{st1,g}>0,\ \widehat X_{st2,g}>0\},
  \qquad
  \rho_g=\frac{K_g^{+}}{K_g},
  \label{eq:support}
\end{align}
where $\rho_g$ is undefined if $K_g=0$.  Define denominator concentration by
\begin{equation}
  c_g
  =
  \max_{s,t,m}
  \frac{\widehat X_{stm,g}}
       {\sum_{s',t',m'}\widehat X_{s't'm',g}}.
  \label{eq:concentration}
\end{equation}
The granularity--stability profile is
\begin{equation}
  \mathcal S_g
  =
  \left(\widehat R_g,
  \widehat{\operatorname{SE}}_g,
  \RSE_g,
  K_g,K_g^{+},\rho_g,c_g\right).
  \label{eq:profile}
\end{equation}

For a nested sequence of domain partitions $\mathcal P_0\preceq\mathcal P_1\preceq\cdots\preceq\mathcal P_L$, the analyst chooses a maximum RSE $\tau$, a minimum paired-support count $K_{\min}$, and, if desired, a maximum concentration $c_{\max}$.  The deepest uniformly reportable level is
\begin{equation}
  \ell^{\star}
  =
  \max\left\{\ell:
  \RSE_g\leq\tau,\ K_g^{+}\geq K_{\min},\ c_g\leq c_{\max}
  \text{ for every }g\in\mathcal P_{\ell}\right\}.
  \label{eq:stopping}
\end{equation}
These thresholds are reporting choices, not universal inferential constants.  The accompanying R script uses transparent defaults that can be changed before analysis.  Reporting the full continuous profile is preferable to reporting only a pass--fail label.

\subsection{Implementation with final and component weights}\label{sec:implementation}

The point estimator can be computed in two equivalent ways when the released final weight has been constructed correctly.  The first sums the final combined weight over the target records.  The second reconstructs \cref{eq:nested-total} from $Z,z,n,H,h,$ and $D^{\star}$.  The code provides both routes.  The reconstruction is valuable as a validation check and as an audit trail; the final-weight route is convenient for repeated tables once the design variables have been verified.  Because the design is with replacement, the reconstruction requires an identifier for sampled first-stage draws rather than merely a unique frame-unit code.

A weight-convention issue must be handled explicitly.  If \texttt{wgt\_combined} is a full-sample weight that already averages the two subsamples, the full point total is the direct weighted sum, while a subsample-specific full-total replicate is obtained by doubling the weighted sum within that subsample.  If the file instead contains subsample-specific weights, the point estimate is the average of the two weighted totals and no doubling is used.  The R script exposes this choice through a single configuration argument and writes the subsample weight totals needed to audit that choice against the file documentation.

\section{Results: India and Himachal Pradesh}\label{sec:results}

\subsection{Analysis status and interpretation}

The four distinct private-institution panels supplied with the project are now incorporated under their original manuscript names: \texttt{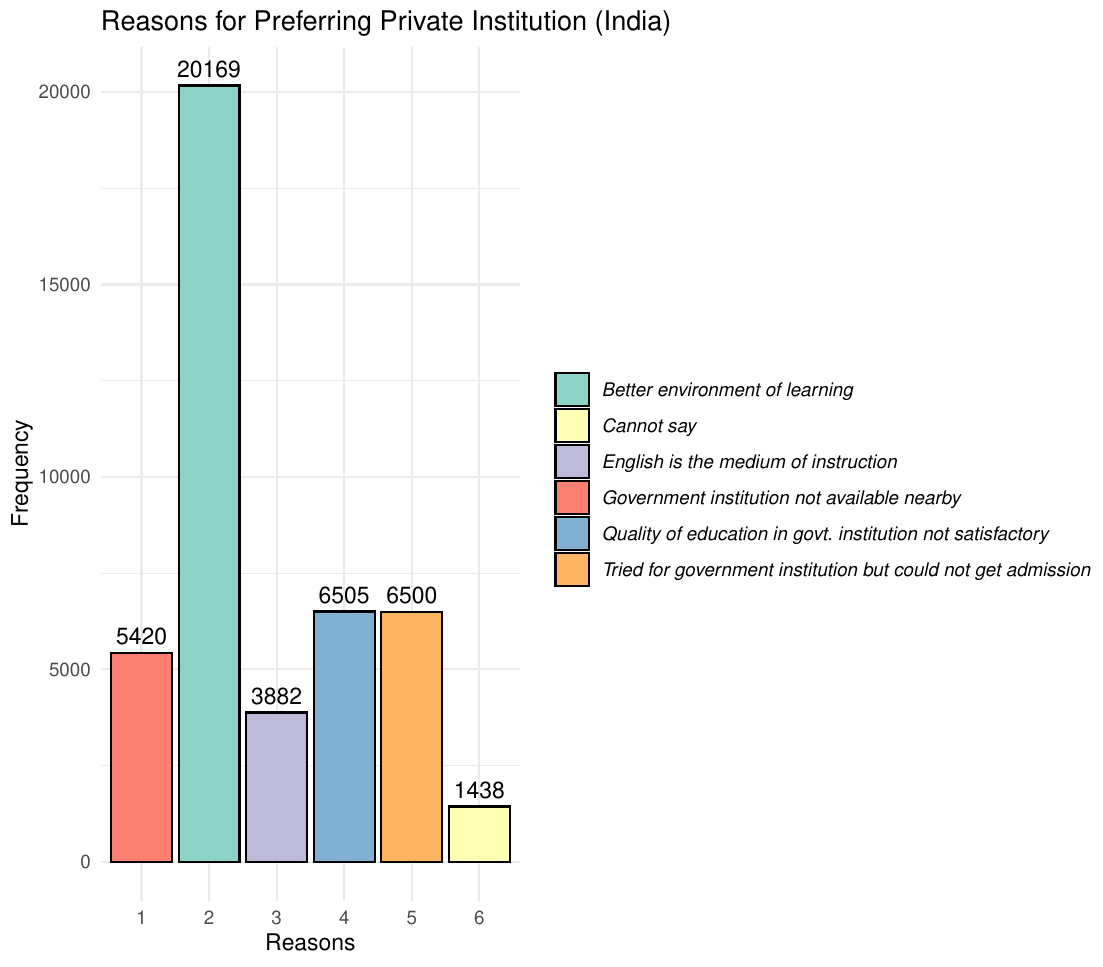} for India, \texttt{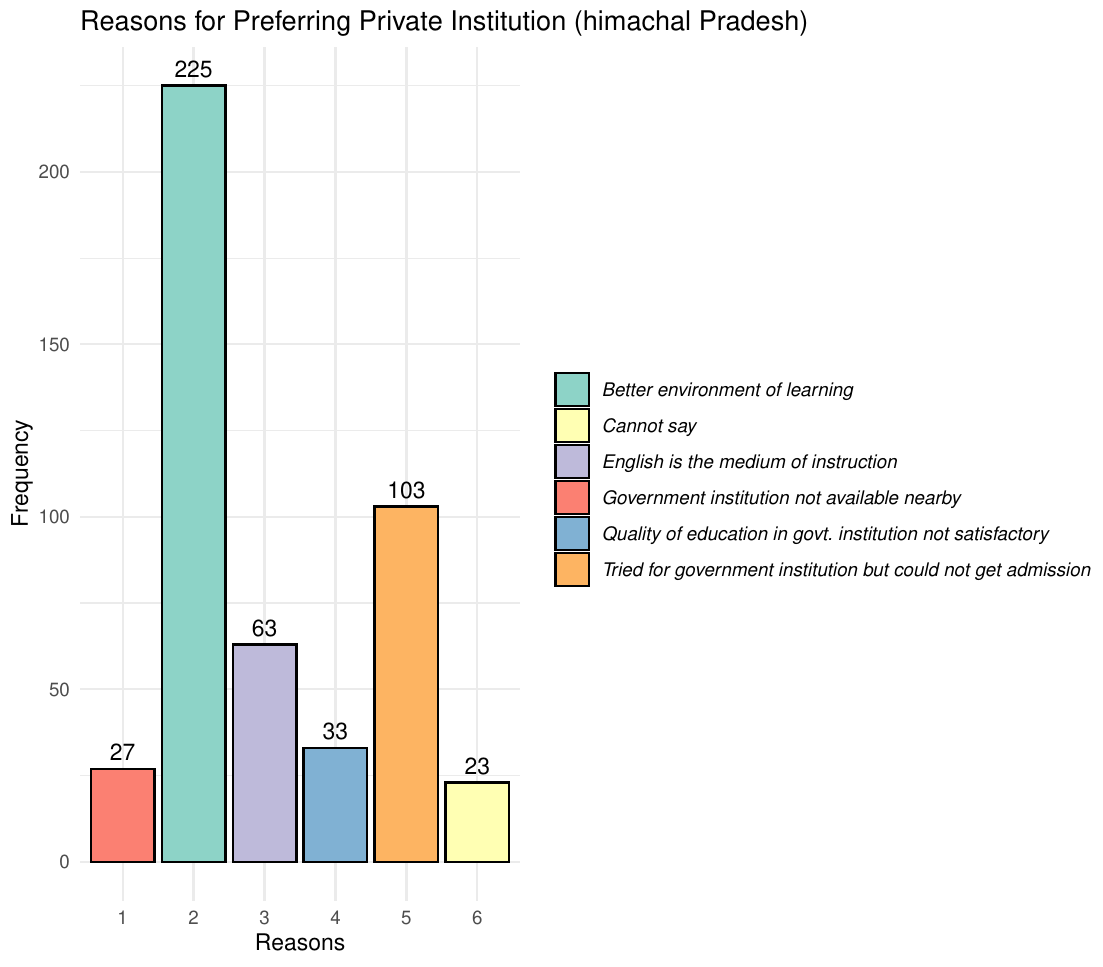} for Himachal Pradesh as a whole, \texttt{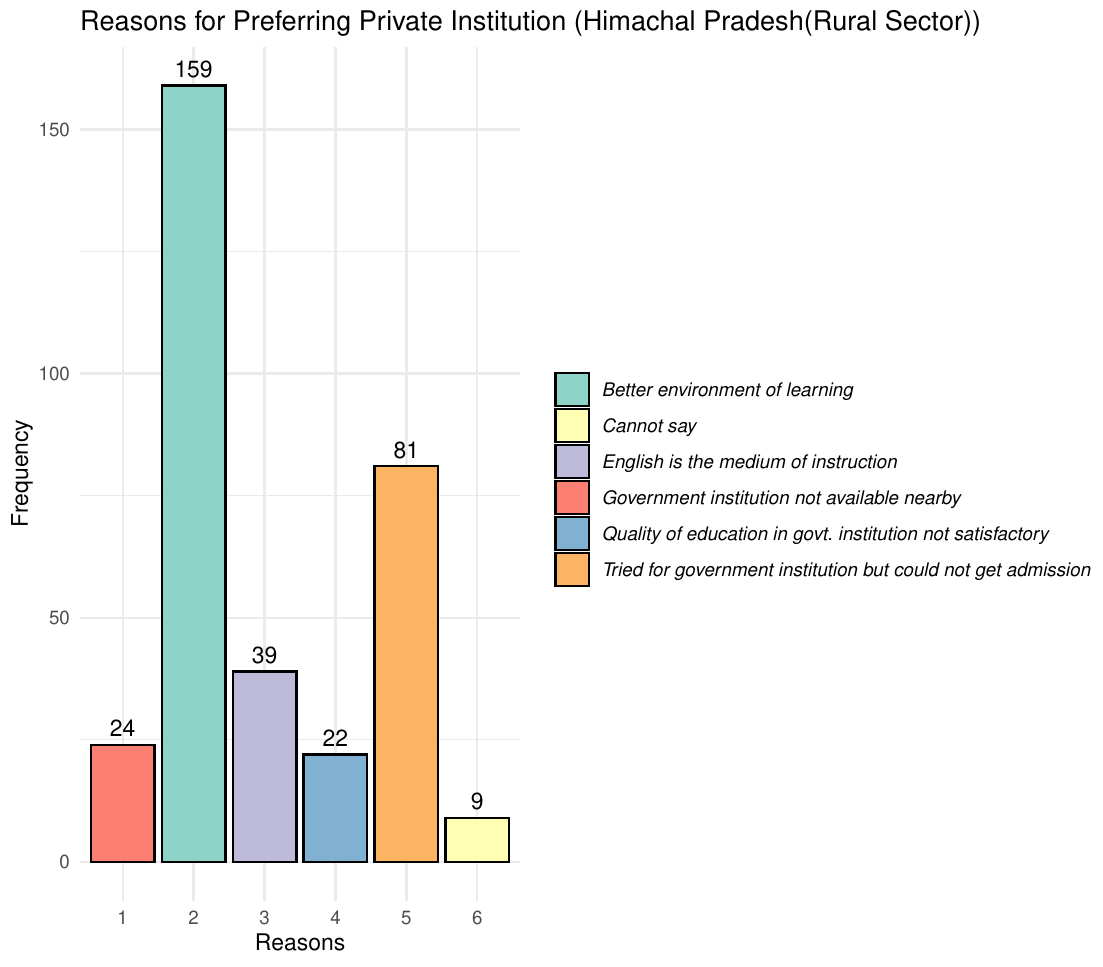} for rural Himachal Pradesh, and \texttt{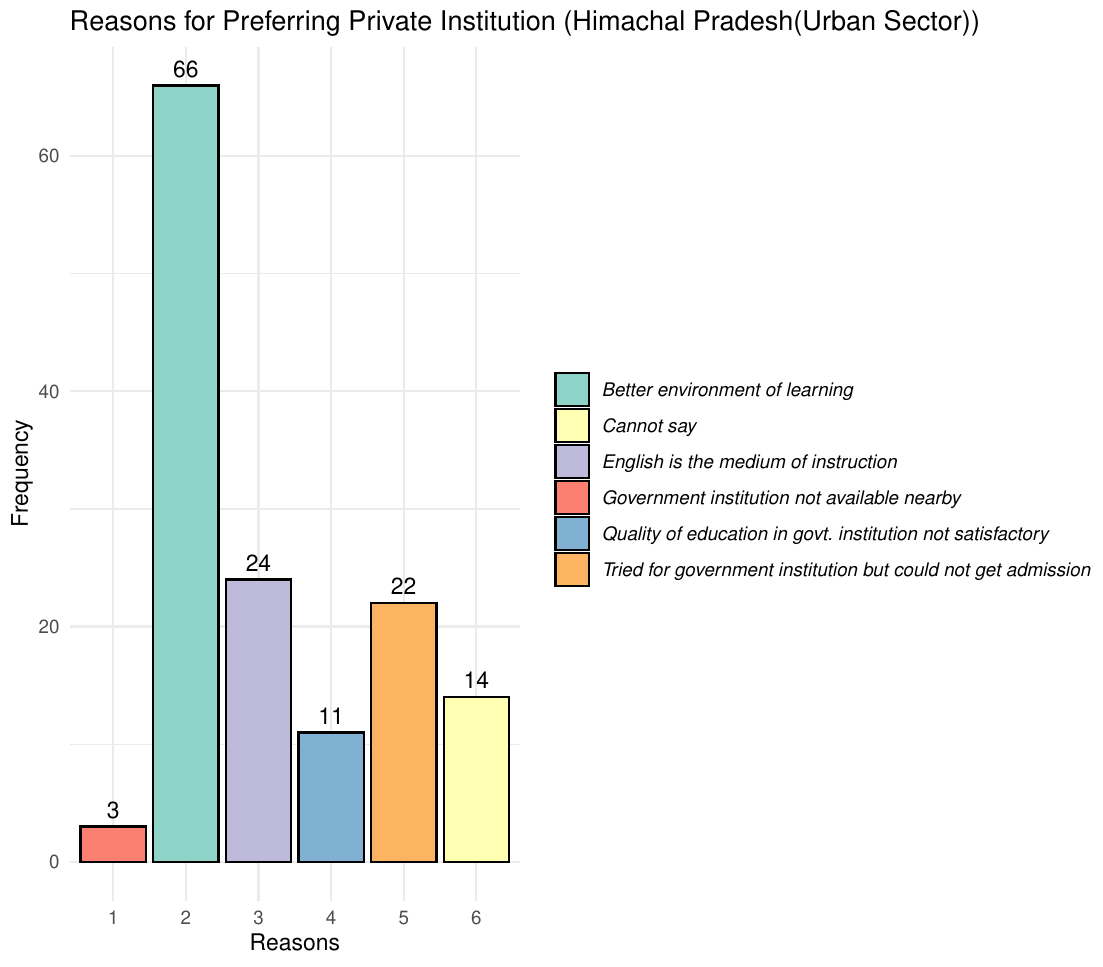} for urban Himachal Pradesh.  The additional file \texttt{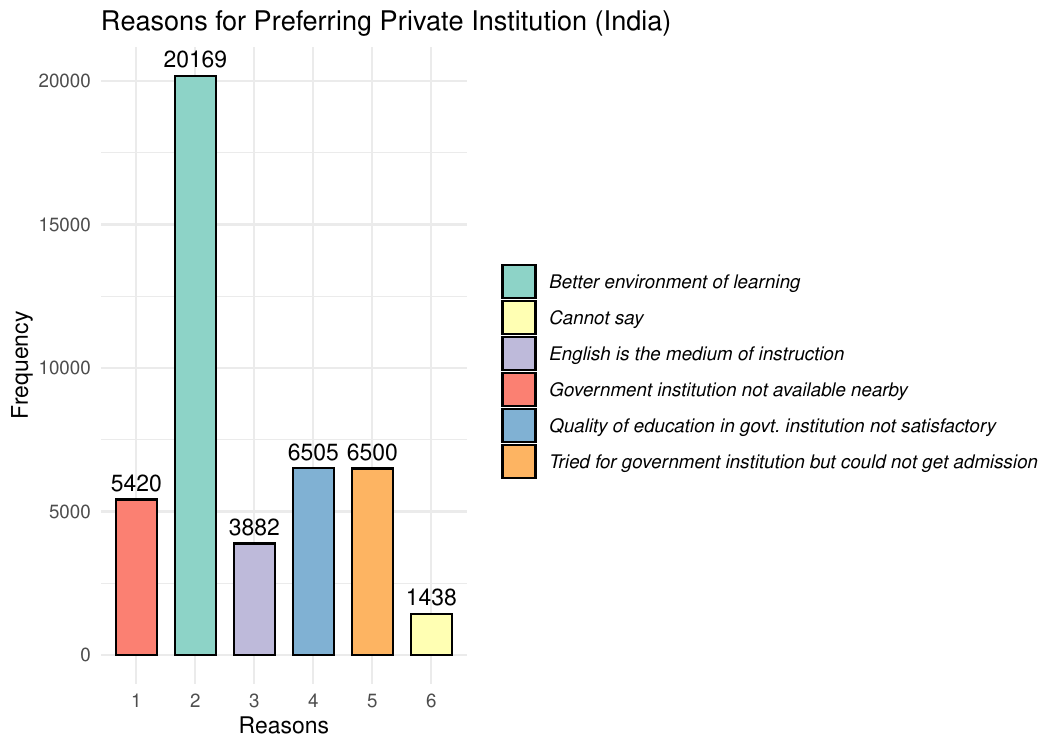} repeats the India panel with a different aspect ratio and is therefore not displayed separately.  The graphics label the vertical axis as ``Frequency,'' print integer counts above the bars, and do not display survey weights or uncertainty intervals.  The numerical discussion below therefore treats them as observed sample frequencies.  Percentages are calculated from the printed counts within each panel and should not be interpreted as design-weighted population estimates.

The underlying microdata and the numerical contents of the 18 externally linked language tables are still not part of the supplied materials.  Consequently, the reason-frequency observations can be verified from the attached figures, whereas the corresponding design-weighted proportions $Q_{c,g}$ in \cref{eq:reason-ratio}, their standard errors, and the granularity--stability diagnostics require regeneration from the public-use files.  The distinction is consequential: the displayed frequencies establish descriptive ordering in the realized sample, but they do not establish population differences or statistical significance.

\subsection{Reported reasons for preferring private educational institutions}\label{sec:private-results}

\begin{figure}[t]
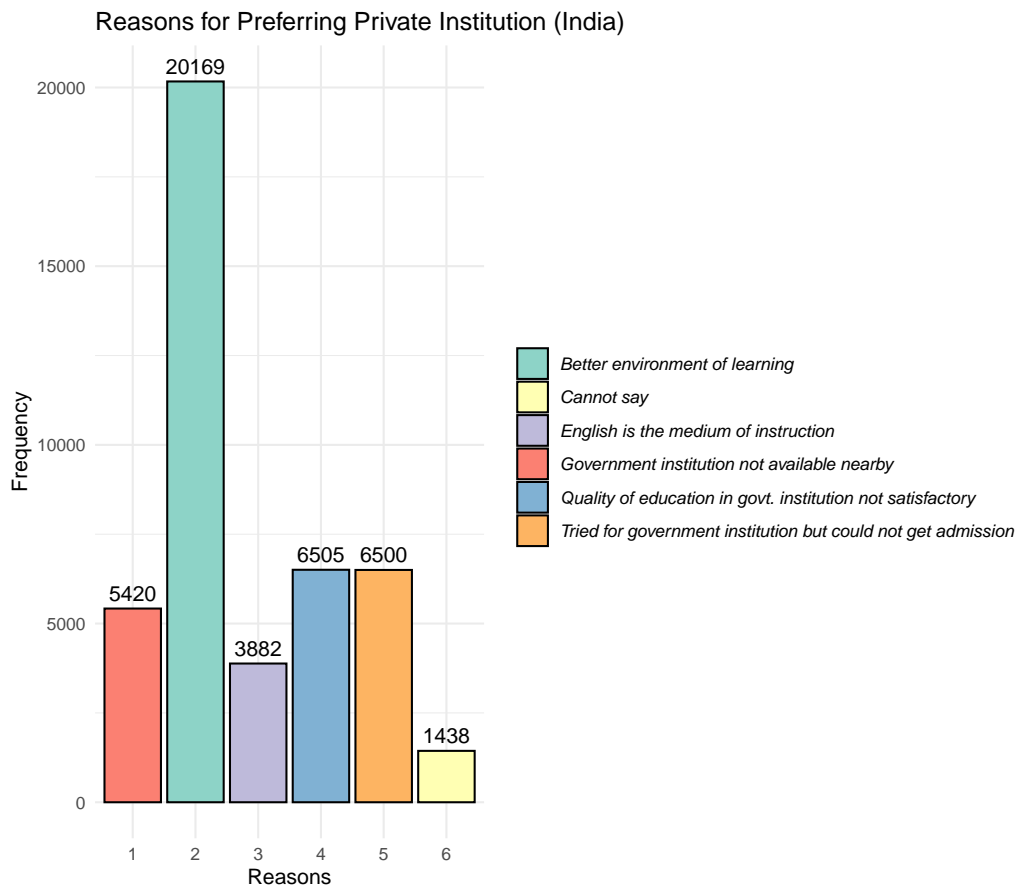

  \centering
  \sourcefigure{Rplot.pdf}{0.82}
  \caption{Displayed sample frequencies for the six reported reasons for preferring a private educational institution in India ($n=43{,}914$).  A better environment of learning is the dominant response, with 20,169 observations (45.9\%).  Unsatisfactory quality in a government institution and an unsuccessful attempt to obtain government admission are nearly tied at 6,505 and 6,500 observations, so the panel supports a clear modal category but not a meaningful ordering of those two adjacent categories without design-based standard errors.}
  \label{fig:india-reasons}
\end{figure}

Across all four panels, bars 1--6 correspond, respectively, to no government institution nearby, a better environment of learning, English as the medium of instruction, unsatisfactory quality in a government institution, an unsuccessful attempt to obtain admission to a government institution, and ``cannot say.''  The India panel in \cref{fig:india-reasons} contains 43,914 recorded responses.  A better environment of learning accounts for 20,169 responses, or 45.9\% of the displayed sample, and is therefore the unambiguous modal category.  Unsatisfactory government-school quality and an unsuccessful attempt to obtain government admission account for 6,505 and 6,500 responses, respectively, both 14.8\%.  Their difference of five observations is too small to support a substantive ranking from the graphic alone.  No nearby government institution contributes 5,420 responses (12.3\%), English-medium instruction 3,882 (8.8\%), and ``cannot say'' 1,438 (3.3\%).

\begin{figure}[t]
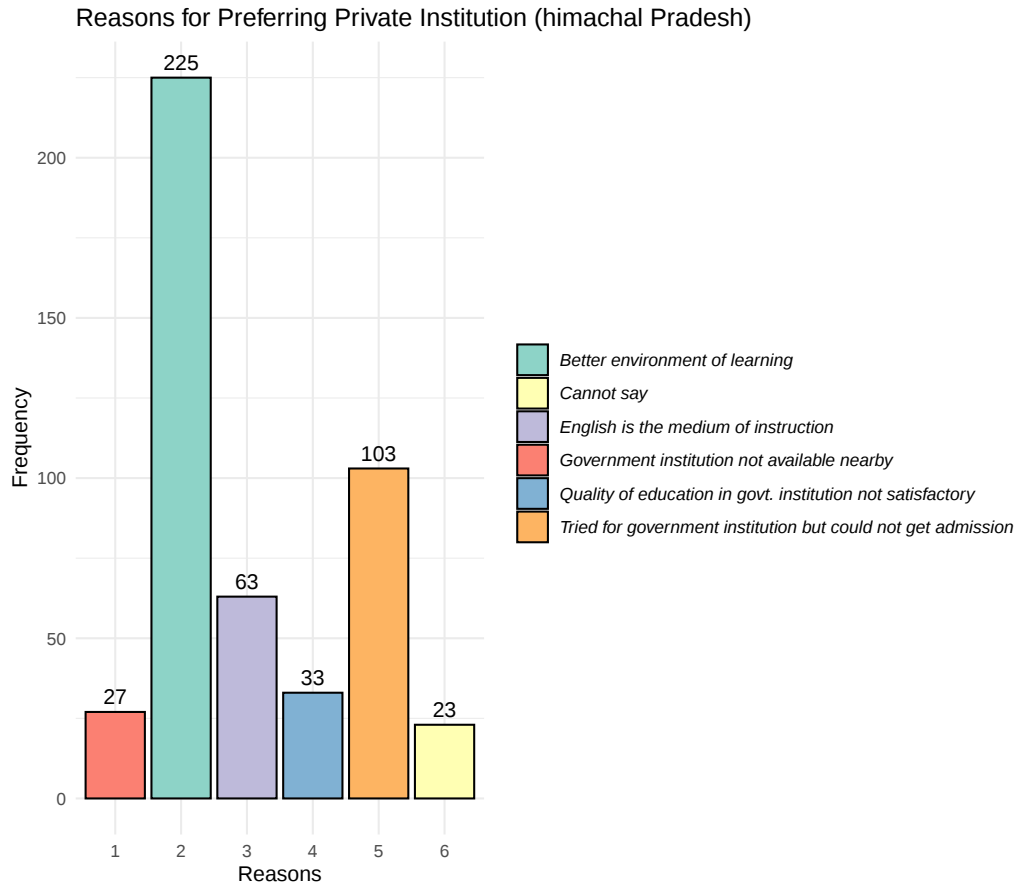

  \centering
  \sourcefigure{Rplot02.pdf}{0.82}
  \caption{Displayed sample frequencies for Himachal Pradesh as a whole ($n=474$).  A better environment of learning remains the modal reason, with 225 observations (47.5\%); an unsuccessful attempt to obtain government admission is second with 103 (21.7\%), and English-medium instruction is third with 63 (13.3\%).  The panel indicates that the quality-oriented modal response persists in the smaller state sample, while admission difficulty forms a substantial secondary component.}
  \label{fig:hp-reasons}
\end{figure}

The Himachal Pradesh panel in \cref{fig:hp-reasons} contains 474 responses.  A better environment of learning again leads, with 225 responses (47.5\%), followed by an unsuccessful attempt to obtain government admission with 103 (21.7\%) and English-medium instruction with 63 (13.3\%).  Unsatisfactory government-school quality, no nearby government institution, and ``cannot say'' contribute 33 (7.0\%), 27 (5.7\%), and 23 (4.9\%) responses.  The displayed sample therefore preserves the national modal category at almost the same share, while placing greater relative mass on unsuccessful government admission and English-medium instruction.  Because the national and state samples arise from different parts of a complex design, this comparison is descriptive and must be reassessed with the weighted ratio estimator before it is stated as a population contrast.

The sectoral panels sharpen this state-level pattern.  Rural Himachal Pradesh contributes 334 responses and urban Himachal Pradesh contributes 140.  The share citing a better environment of learning is almost unchanged across sectors: 159 of 334 rural responses (47.6\%) and 66 of 140 urban responses (47.1\%).  The remaining composition differs more visibly.  An unsuccessful attempt to obtain government admission accounts for 24.3\% in the rural sample and 15.7\% in the urban sample, while the absence of a nearby government institution accounts for 7.2\% and 2.1\%, respectively.  Conversely, English-medium instruction is cited by 11.7\% of rural and 17.1\% of urban responses, and ``cannot say'' by 2.7\% and 10.0\%.  Unsatisfactory government-school quality is comparatively similar at 6.6\% in rural areas and 7.9\% in urban areas.  Thus the images refine the original qualitative conclusion: admission and proximity constraints are more visible in the rural sample, whereas English-medium instruction and noncommittal responses occupy larger shares in the urban sample.

\begin{figure}[t]
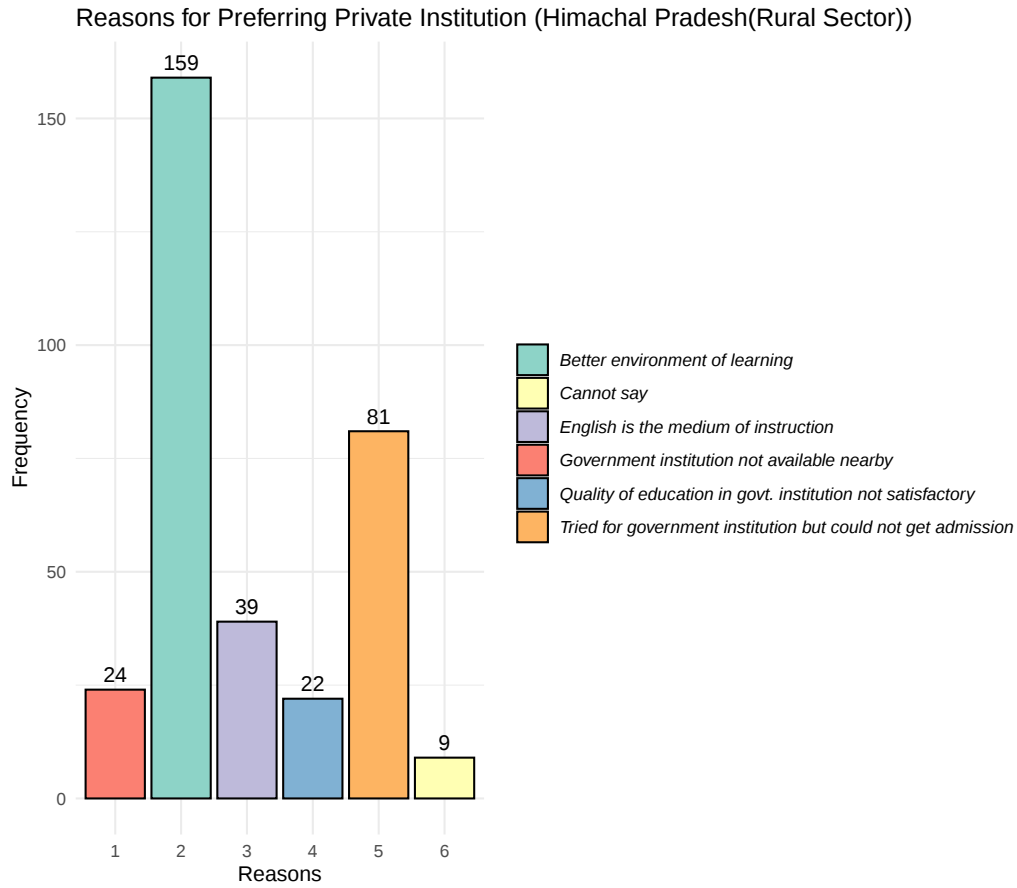

  \centering
  \sourcefigure{Rplot03.pdf}{0.82}
  \caption{Displayed sample frequencies for rural Himachal Pradesh ($n=334$).  A better environment of learning accounts for 159 observations (47.6\%), followed by an unsuccessful attempt to obtain government admission with 81 (24.3\%).  No nearby government institution is cited 24 times (7.2\%).  Relative to the urban panel, the rural composition assigns more of its sample frequency to admission and proximity constraints, although population inference requires the design-weighted standard errors.}
  \label{fig:hp-rural-reasons}
\end{figure}

\begin{figure}[t]
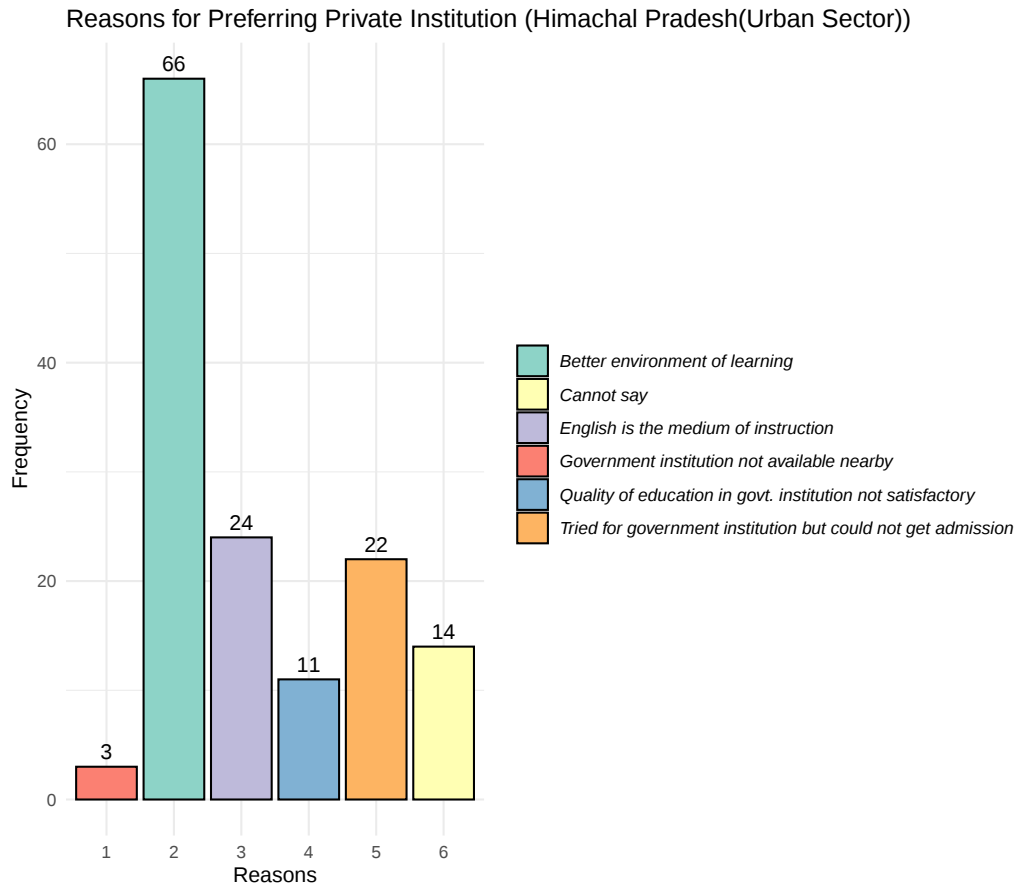

  \centering
  \sourcefigure{Rplot04.pdf}{0.82}
  \caption{Displayed sample frequencies for urban Himachal Pradesh ($n=140$).  A better environment of learning remains modal with 66 observations (47.1\%); English-medium instruction and an unsuccessful attempt to obtain government admission contribute 24 (17.1\%) and 22 (15.7\%), while ``cannot say'' contributes 14 (10.0\%).  Only three responses (2.1\%) cite the absence of a nearby government institution.  The panel preserves the statewide quality-oriented modal category but shows a more English-oriented and less proximity-constrained composition than the rural sample.}
  \label{fig:hp-urban-reasons}
\end{figure}

\FloatBarrier

The state and sector panels also provide an exact computational coherence check.  In the category order stated above,
\begin{equation}
  (27,225,63,33,103,23)
  =
  (24,159,39,22,81,9)
  +(3,66,24,11,22,14).
  \label{eq:observed-sector-reconciliation}
\end{equation}
Hence every Himachal Pradesh count equals the corresponding rural count plus the urban count, and $474=334+140$ for the panel totals.  This finite-sample reconciliation is concordant with the exact domain-partition identity in \cref{prop:domain-additivity}: the displayed counts are internally coherent under the rural--urban partition, and a design-weighted reconstruction should satisfy the same algebraic identity.  The concordance is an accounting check rather than an empirical verification of design unbiasedness in \cref{thm:unbiased}.

The bars correspond substantively to category proportions, so population inference is governed by the ratio representation in \cref{thm:ratio} and the variance result in \cref{cor:ratio-var}.  Those results do not predict which reason should be largest; they state how uncertainty must account for both a category total and the estimated number of eligible private-institution respondents.  Because the figures contain neither weights nor subsample differences, apparent gaps should not be interpreted as statistically significant.  The India counts of 6,505 and 6,500 illustrate the point especially clearly, and the rural--urban differences in Himachal Pradesh require the same design-based qualification.

Finally, the successive panel sizes---43,914 for India, 474 for Himachal Pradesh, 334 for rural Himachal Pradesh, and 140 for urban Himachal Pradesh---make the paper's granularity problem visible.  The modal better-environment category remains close to 46--48\% throughout the displayed samples, but several sector-specific cells become very small; the urban proximity category contains only three observations.  The empirical pattern therefore supports the broad methodological argument without overstating it: stable qualitative ordering can coexist with rapidly diminishing information for secondary categories, which is precisely why RSE, paired-subsample support, and denominator concentration must accompany deeply disaggregated estimates.

The findings remain descriptive rather than causal.  They do not imply that attendance at a private institution improves learning, nor do they identify the effect of distance, admission constraints, or instructional language on institution choice.

\subsection{Home language and medium of instruction}\label{sec:language-results}

The language analysis compares weighted distributions of the language mainly spoken at home and the medium of instruction over age, sector, sex, and geography.  The source project reports a larger English-medium share in urban Himachal Pradesh than in rural Himachal Pradesh.  In the transition-table formulation of \cref{eq:joint-language}, this statement concerns the urban--rural contrast in the English column and, more specifically, the off-diagonal mass contributed by students whose home language is not English.  The distinction matters because a high English-medium share can arise either from a high English home-language share or from a transition into English at school.  The joint table makes those mechanisms visible.

The sex-specific tables extend the comparison without changing the estimand.  For each domain, row sums describe the home-language composition, column sums describe the medium-of-instruction composition, and diagonal mass contributes to the agreement rate in \cref{eq:match-rate}.  The RSE attached to every cell determines whether a detailed transition is reportable.  A rare language can have a substantively interesting point estimate and nevertheless lack enough paired first-stage support for a stable direct estimate.  This is precisely the setting in which the granularity--stability profile is more informative than an unqualified percentage table.

The original project stored the estimates, estimated standard deviations, and coefficients of variation in external tables.  Those links are retained in \cref{tab:hp-language-links,tab:india-language-links}.  The revised R script writes a single machine-readable table containing the estimate, standard error, CV/RSE, active design cells, paired support, and concentration for every language-domain-age combination.  The linked tables can therefore remain as presentation files while the combined output serves as the reproducible statistical record.

\begin{table}[htbp]
\centering
\begin{threeparttable}
\caption{Reported language-estimation tables for Himachal Pradesh.}
\label{tab:hp-language-links}
\begin{tabularx}{0.93\textwidth}{@{}Xl@{}}
\toprule
Domain & Appendix Reference \\
\midrule
Himachal Pradesh, female & Table \ref{tab:csv_hp_all_female} \\
Himachal Pradesh, male & Table \ref{tab:csv_hp_all_male} \\
Himachal Pradesh, all persons & Table \ref{tab:csv_hp_all_neutral} \\
Himachal Pradesh, rural female & Table \ref{tab:csv_hp_rural_female} \\
Himachal Pradesh, rural male & Table \ref{tab:csv_hp_rural_male} \\
Himachal Pradesh, rural, all persons & Table \ref{tab:csv_hp_rural_neutral} \\
Himachal Pradesh, urban female & Table \ref{tab:csv_hp_urban_female} \\
Himachal Pradesh, urban male & Table \ref{tab:csv_hp_urban_male} \\
Himachal Pradesh, urban, all persons & Table \ref{tab:csv_hp_urban_neutral} \\
\bottomrule
\end{tabularx}
\begin{tablenotes}[flushleft]
\footnotesize
\item The source draft describes these files as tables of estimated percentages with estimated standard deviations and coefficients of variation. Their numerical contents were not embedded in the supplied manuscript source, but are visually rendered in Appendix \ref{sec:appendix_data_tables}.
\end{tablenotes}
\end{threeparttable}
\end{table}

\begin{table}[htbp]
\centering
\begin{threeparttable}
\caption{Reported language-estimation tables for India.}
\label{tab:india-language-links}
\begin{tabularx}{0.93\textwidth}{@{}Xl@{}}
\toprule
Domain & Appendix Reference \\
\midrule
India, female & Table \ref{tab:csv_ind_all_female} \\
India, male & Table \ref{tab:csv_ind_all_male} \\
India, all persons & Table \ref{tab:csv_ind_all_neutral} \\
India, rural female & Table \ref{tab:csv_ind_rural_female} \\
India, rural male & Table \ref{tab:csv_ind_rural_male} \\
India, rural, all persons & Table \ref{tab:csv_ind_rural_neutral} \\
India, urban female & Table \ref{tab:csv_ind_urban_female} \\
India, urban male & Table \ref{tab:csv_ind_urban_male} \\
India, urban, all persons & Table \ref{tab:csv_ind_urban_neutral} \\
\bottomrule
\end{tabularx}
\begin{tablenotes}[flushleft]
\footnotesize
\item The accompanying code produces a consolidated dataset visually rendered in Appendix \ref{sec:appendix_data_tables}.
\end{tablenotes}
\end{threeparttable}
\end{table}

\FloatBarrier

\subsection{What the empirical analysis contributes to the methodological argument}\label{sec:results-method-link}

The private-institution and language analyses occupy different parts of the outcome space.  The reason variable has only six categories, yet the attached panels show how quickly some categories become sparse after geographic and sectoral splitting.  The India panel contains tens of thousands of observations, whereas the entire urban Himachal Pradesh panel contains 140 and one reason category contains only three.  The language transition table can be sparse for a different reason: it contains many home-language--instruction-language cells even before sex and age are introduced.  Together, the two analyses demonstrate the paper's two routes to instability---a small domain denominator and a large number of categories within that denominator.

The better-environment category is descriptively robust across the displayed samples, accounting for 45.9\% in India, 47.5\% in Himachal Pradesh, 47.6\% in rural Himachal Pradesh, and 47.1\% in urban Himachal Pradesh.  This persistence is a substantive observation, not a theorem-driven consequence.  By contrast, the exact rural-plus-urban reconstruction of every state count is a deterministic coherence property and is directly concordant with \cref{prop:domain-additivity}.  The distinction matters: algebraic reconciliation can be verified from the figures, whereas stability of the corresponding population proportions requires the linearization and replication calculations in \cref{thm:ratio,thm:replication,cor:ratio-var}.

The framework therefore preserves the substantive question at every level while changing the evidential status of the answer according to the realized design information.  A dominant category with broad support can be reported with its design-based uncertainty; a rare urban or language-transition cell may require aggregation or explicit qualification.  This avoids the two common extremes of suppressing all detailed analysis and reporting every percentage as if its precision were comparable.

\begin{table}[t]
\centering
\caption{Observed figure-based findings and their inferential role.}
\label{tab:reported-findings}
\begin{tabularx}{0.98\textwidth}{@{}>{\raggedright\arraybackslash}p{0.22\textwidth}>{\raggedright\arraybackslash}X>{\raggedright\arraybackslash}X@{}}
\toprule
Comparison & Observation from the supplied material & Role in the present paper \\
\midrule
India, private-institution reasons & Better learning environment: 20,169 of 43,914 displayed responses (45.9\%); the two 14.8\% categories differ by only five observations. & Establishes the national sample ordering and illustrates why visually adjacent bars should not be ranked without design-based uncertainty. \\
Himachal Pradesh, all sectors & Better learning environment: 225 of 474 (47.5\%); unsuccessful government admission: 103 (21.7\%). & Shows persistence of the modal category in a much smaller state sample while exposing a different secondary composition. \\
Himachal Pradesh, rural versus urban & The better-environment share is nearly identical (47.6\% versus 47.1\%); admission and proximity constraints are more frequent in the rural sample, whereas English-medium and ``cannot say'' are more frequent in the urban sample.  All six state counts equal rural plus urban counts exactly. & Motivates sector-specific precision diagnostics and supplies an empirical accounting check concordant with \cref{prop:domain-additivity}. \\
Himachal Pradesh, instructional medium & The source project reports a larger English-medium share in urban than rural areas. & Motivates the home-to-school transition table and sector-specific RSE diagnostics; the numerical language tables remain external. \\
\bottomrule
\end{tabularx}
\end{table}

\FloatBarrier

\section{Conclusion}\label{sec:conclusion}

This paper recasts an education-survey project as a design-based methodological analysis of inferential granularity.  The central object is a nested contribution ledger that follows the NSS estimator through household expansion, hamlet-group or sub-block selection, first-stage probability-proportional-to-size expansion, and two-subsample aggregation.  The ledger supports arbitrary domain totals without altering the weights, and smooth domain parameters are obtained as ratios of those totals.  Linearization and the two independent subsamples yield a common variance calculation across all estimands.

The resulting granularity--stability profile makes the paper's contribution more specific than a generic recommendation to ``account for survey design.''  It records not only RSE but also whether both subsamples support the domain and whether the estimated denominator is concentrated in a few design cells.  Applied over a nested sequence of partitions, the profile identifies the deepest level at which direct design-based reporting remains credible.  Domains beyond that level are not discarded silently: they are marked as requiring aggregation, qualification, or a separate model-based small-area analysis.

The attached frequency panels sharpen the private-institution findings.  A better learning environment accounts for 45.9\% of displayed responses in India and approximately 47\% in Himachal Pradesh overall and in each sector.  Within the state sample, unsuccessful government admission and the absence of a nearby government institution are more prominent in rural areas, whereas English-medium instruction and ``cannot say'' occupy larger shares in urban areas.  The state counts reconcile exactly with their rural and urban components, as required by \cref{prop:domain-additivity}.  These percentages are computed from the printed sample frequencies; the statistical framework determines how strongly the corresponding population statements are supported once survey weights, subsample variation, and the relevant geographic, sectoral, sex, age, and language classifications are imposed.  The source project also reports a larger English-medium share among urban than rural students in Himachal Pradesh, but the numerical language tables remain external.

Several limitations are deliberate.  The analysis is descriptive and design-based; it does not estimate causal effects of institution type or instructional language.  It does not correct residual nonresponse bias beyond the released weighting system.  It also does not borrow strength across unstable domains.  Empirical best linear unbiased prediction and Fay--Herriot models \citep{fay1979}, calibration to Census totals, and design-adapted bootstrap validation are useful extensions when the required auxiliary data and modeling assumptions are available.  They should be introduced after, rather than instead of, a transparent assessment of the direct estimator.

Finally, the original figure files are now incorporated and their printed counts have been checked against one another, including exact rural--urban reconciliation for Himachal Pradesh.  The microdata and numerical language tables were not supplied, so the displayed frequency summaries could not be converted here into design-weighted reason proportions or independently recomputed language estimates.  The separate R file resolves the duplicated and malformed portions of the original script, uses label-aware state and language handling, implements the correct ratio linearization, reproduces the sample-frequency panels under the corrected file mapping, and writes weighted estimates and diagnostics in reproducible form.  Re-running that script on the public-use files is the remaining empirical step before submission.

\appendix

\section{Proofs}\label{app:proofs}

\subsection{Proof of design unbiasedness}

Fix a stratum $s$, sub-stratum $t$, subsample $m$, and second-stage stratum $j$.  Suppress these indices when no ambiguity arises.  Let
\begin{equation}
  T_{idj}=\sum_{k=1}^{H_{idj}}y_{idjk}
\end{equation}
be the finite-population total in hamlet-group or sub-block $d$ of first-stage unit $i$.  Under simple random sampling without replacement of $h_{idj}$ households from $H_{idj}$,
\begin{equation}
  \E_p\left(
  \frac{H_{idj}}{h_{idj}}
  \sum_{k\in s_{idj}}y_{idjk}
  \ \middle|\ i,d,j
  \right)
  =T_{idj}.
  \label{eq:proof-household}
\end{equation}

If $D_i=1$, the first selected group is the whole first-stage unit, and \cref{eq:proof-household} gives conditional unbiasedness for its total.  Suppose $D_i>1$.  Group 1 is selected with certainty and a random index $J_i$ is drawn uniformly from $\{2,\ldots,D_i\}$.  After taking expectation over household selection, the intermediate-stage estimator has conditional expectation
\begin{align}
  \E_p\left\{T_{i1j}+(D_i-1)T_{iJ_i j}\mid i\right\}
  &
  =T_{i1j}
  +(D_i-1)\frac{1}{D_i-1}
  \sum_{d=2}^{D_i}T_{idj}
  \nonumber\\
  &=\sum_{d=1}^{D_i}T_{idj}
  \equiv T_{ij}.
  \label{eq:proof-hamlet}
\end{align}
Thus the expression in braces in \cref{eq:nested-total} is conditionally unbiased for the second-stage-stratum total $T_{ij}$ in first-stage unit $i$.

Let $I_1,\ldots,I_n$ denote the independent probability-proportional-to-size draws in the sub-stratum, with $\Pr(I_r=i)=p_i=z_i/Z$.  Let $\widetilde T_{I_rj}$ denote the conditionally unbiased estimator from \cref{eq:proof-hamlet}.  Iterated expectation gives
\begin{align}
  \E_p\left\{
  \frac{Z}{n}\sum_{r=1}^{n}\frac{\widetilde T_{I_rj}}{z_{I_r}}
  \right\}
  &=
  \frac{Z}{n}\sum_{r=1}^{n}
  \sum_i\frac{z_i}{Z}\frac{T_{ij}}{z_i}
  \\
  &=\sum_i T_{ij}.
  \label{eq:proof-pps}
\end{align}
Summing \cref{eq:proof-pps} over $j$ gives the finite-population total for the sub-stratum.  Each of the two subsample estimators is unbiased for that same total, so their average is unbiased.  Summing over strata and sub-strata completes the proof.  The argument remains unchanged when $y$ is multiplied by a fixed domain indicator. \qed

\subsection{Proof of exact domain additivity}

For every finite-population unit, mutual exclusivity and $G=\sum_{r=1}^q G_r$ imply the pointwise identity
\begin{equation}
  G y=\sum_{r=1}^q G_r y.
\end{equation}
Every operation in \cref{eq:nested-total,eq:aggregation} is linear in the supplied characteristic: household values are summed, multiplied by design constants, and then summed or averaged over sampling stages.  Applying that linearity to the pointwise identity gives
\begin{equation}
  \widehat T(Gy)
  =\widehat T\!\left(\sum_{r=1}^q G_r y\right)
  =\sum_{r=1}^q\widehat T(G_r y),
\end{equation}
for every realized sample.  Taking $q=2$ with rural and urban indicators gives the stated sector decomposition. \qed

\subsection{Proof of the first-order ratio representation}

Write $\widehat Y=\widehat T_{\nu}(y)$, $Y=T_{\nu}(y)$, $\widehat X=\widehat T_{\nu}(x)$, $X=T_{\nu}(x)$, and $R=Y/X$.  The ratio error has the exact representation
\begin{equation}
  \widehat R-R
  =
  \frac{\widehat Y}{\widehat X}-\frac{Y}{X}
  =
  \frac{(\widehat Y-Y)-R(\widehat X-X)}{\widehat X}.
  \label{eq:proof-ratio-exact}
\end{equation}
By \cref{ass:ratio}, $\widehat X/X\xrightarrow{p}1$, and hence
\begin{equation}
  \frac{1}{\widehat X}
  =
  \frac{1}{X}\{1+o_p(1)\}.
\end{equation}
Multiplying \cref{eq:proof-ratio-exact} by $a_{\nu}^{1/2}$ and using the tightness implied by \cref{eq:joint-clt} yields \cref{eq:ratio-linearization}.  If $R_{\nu}\to R_0$, the continuous mapping theorem and Slutsky's theorem give
\begin{equation}
  a_{\nu}^{1/2}(\widehat R_{\nu}-R_{\nu})
  \xrightarrow{d}
  N\left(0,
  \begin{pmatrix}1&-R_0\end{pmatrix}
  \bm\Sigma
  \begin{pmatrix}1\\-R_0\end{pmatrix}
  \right).
\end{equation}
The variance approximation in \cref{eq:ratio-variance-linearized} is the variance of the leading term in \cref{eq:ratio-linearization}. \qed

\subsection{Proof of the two-subsample variance result}

Fix $(s,t)$ and abbreviate $\widehat T_{st1}(\ell)$ and $\widehat T_{st2}(\ell)$ by $L_1$ and $L_2$.  By assumption, $\E_p(L_1)=\E_p(L_2)=L$, $\Var_p(L_1)=\Var_p(L_2)=V$, and $\Cov_p(L_1,L_2)=0$.  Therefore
\begin{align}
  \E_p\{(L_1-L_2)^2\}
  &=\Var_p(L_1-L_2)
    +\{\E_p(L_1-L_2)\}^2
  \\
  &=2V,
\end{align}
and
\begin{equation}
  \Var_p\left(\frac{L_1+L_2}{2}\right)
  =\frac{1}{4}\{\Var_p(L_1)+\Var_p(L_2)\}
  =\frac{V}{2}.
\end{equation}
It follows that
\begin{equation}
  \E_p\left\{\frac{1}{4}(L_1-L_2)^2\right\}
  =\frac{V}{2}
  =\Var_p\left(\frac{L_1+L_2}{2}\right).
\end{equation}
Independence across stratum--sub-stratum pairs permits summation of both variances and unbiased variance estimators, proving \cref{eq:linear-var}. \qed

\subsection{Proof of the ratio-variance corollary}

Let $\ell_{\nu}=y-R_{\nu}x$ and abbreviate
\begin{equation}
  S_{\ell,\nu}=\sum_{s,t}D_{st,\nu}(\ell_{\nu})^2,
  \qquad
  S_{x,\nu}=\sum_{s,t}D_{st,\nu}(x)^2,
  \qquad
  A_{\nu}=\frac{1}{4}S_{\ell,\nu}.
\end{equation}
By \cref{thm:replication}, $\E_p(A_{\nu})=V_{\nu}$.  Distinct stratum--sub-stratum cells are independent under \cref{ass:subsamples}, so
\begin{equation}
  \Var_p(A_{\nu})
  =\frac{1}{16}\sum_{s,t}
    \Var_p\{D_{st,\nu}(\ell_{\nu})^2\}.
\end{equation}
The first part of \cref{eq:rep-regularity} and Chebyshev's inequality therefore imply $A_{\nu}/V_{\nu}\xrightarrow{p}1$.

Write $\delta_{\nu}=\widehat R_{\nu}-R_{\nu}$ and let
\begin{equation}
  B_{\nu}
  =\frac{1}{4}\sum_{s,t}
   \{D_{st,\nu}(\ell_{\nu})-\delta_{\nu}D_{st,\nu}(x)\}^2
\end{equation}
be the numerator of the plug-in replication estimator.  By the Cauchy--Schwarz inequality,
\begin{align}
  \frac{|B_{\nu}-A_{\nu}|}{V_{\nu}}
  &\leq
  \frac{|\delta_{\nu}|}{2V_{\nu}}
  \{S_{\ell,\nu}S_{x,\nu}\}^{1/2}
  +\frac{\delta_{\nu}^2S_{x,\nu}}{4V_{\nu}}
  \\
  &=
  \frac{1}{2}
  \left\{\frac{\delta_{\nu}^2S_{x,\nu}}{V_{\nu}}
           \frac{S_{\ell,\nu}}{V_{\nu}}\right\}^{1/2}
  +\frac{1}{4}\frac{\delta_{\nu}^2S_{x,\nu}}{V_{\nu}}
  \xrightarrow{p}0,
\end{align}
where the last step uses $S_{\ell,\nu}/V_{\nu}=4A_{\nu}/V_{\nu}\xrightarrow{p}4$ and the second part of \cref{eq:rep-regularity}.  Hence $B_{\nu}/V_{\nu}\xrightarrow{p}1$.  Finally, \cref{ass:ratio} gives $T_{\nu}(x)^2/\widehat T_{\nu}(x)^2\xrightarrow{p}1$, and therefore
\begin{equation}
  \frac{\widehat V(\widehat R_{\nu})}
       {V_{\nu}/T_{\nu}(x)^2}
  =
  \frac{B_{\nu}}{V_{\nu}}
  \frac{T_{\nu}(x)^2}{\widehat T_{\nu}(x)^2}
  \xrightarrow{p}1.
\end{equation}
This proves \cref{eq:ratio-var-consistency}. \qed

\section{Survey-design details retained from the original project}\label{app:design}

\subsection{Frame, stratification, and allocation}

The rural first-stage frame was the 2011 Census village list, except that the available 2001-based Panchayat-ward list was used for Kerala.  The urban frame was the updated Urban Frame Survey block list for phase 2007--2012.  A district was generally divided into a rural stratum and an urban stratum.  Within urban areas, each town with population at least 100,000 formed a separate basic stratum, and the remaining urban area formed another stratum.

If a rural stratum received $r$ sampled first-stage units, it was divided into $r/2$ sub-strata after sorting villages by population, with approximately equal population in each sub-stratum.  If an urban stratum received $u$ sampled units, it was divided into $u/2$ sub-strata after sorting blocks by number of households, with approximately equal household counts.  Two first-stage units were allocated to each sub-stratum.

The central sample contained 8,300 first-stage units at the all-India level.  State and union-territory allocations were proportional to Census 2011 population subject to minimum allocations and field-resource constraints.  State-level allocations were split between rural and urban sectors in proportion to population with double weight for the urban sector, subject to restrictions for large states.  At least 16 first-stage units---eight rural and eight urban---were allocated to each state or union territory.  Stratum allocations were proportional to population, adjusted to even numbers with a minimum of two first-stage units.  The source project also mentioned 9,274 first-stage units for a state sample; that number is not needed for the present central-sample analysis and should be checked against the separate state-sample documentation if it is to be reported.

Within rural sub-strata, villages were selected by probability proportional to size with replacement, using Census population as size.  Within urban sub-strata, Urban Frame Survey blocks were selected by probability proportional to size with replacement, using number of households as size.  The rural and urban samples were each drawn as two independent subsamples.

\subsection{Hamlet-group and sub-block formation}

For most areas, the population thresholds for subdividing a sampled first-stage unit were as follows.

\begin{table}[h]
\centering
\caption{General population rule for forming hamlet-groups or sub-blocks.}
\label{tab:hg-general}
\begin{tabularx}{0.96\textwidth}{@{}>{\raggedright\arraybackslash}X>{\raggedleft\arraybackslash}p{0.40\textwidth}@{}}
\toprule
Approximate present population of sampled first-stage unit & Number of groups \\
\midrule
Less than 1,200 & 1 \\
1,200--1,799 & 3 \\
1,800--2,399 & 4 \\
2,400--2,999 & 5 \\
3,000--3,599 & 6 \\
Thereafter & One additional group per additional 600 persons \\
\bottomrule
\end{tabularx}
\end{table}

A lower threshold applied to rural areas of Himachal Pradesh, Sikkim, most of Uttarakhand, selected districts of Jammu and Kashmir, and Idukki district of Kerala.

\begin{table}[h]
\centering
\caption{Special rural population rule applicable to Himachal Pradesh and other designated areas.}
\label{tab:hg-special}
\begin{tabularx}{0.96\textwidth}{@{}>{\raggedright\arraybackslash}X>{\raggedleft\arraybackslash}p{0.42\textwidth}@{}}
\toprule
Approximate present population of sampled village & Number of hamlet-groups \\
\midrule
Less than 600 & 1 \\
600--899 & 3 \\
900--1,199 & 4 \\
1,200--1,499 & 5 \\
1,500--1,799 & 6 \\
Thereafter & One additional group per additional 300 persons \\
\bottomrule
\end{tabularx}
\end{table}

Groups were formed to have approximately equal population and identifiable physical boundaries.  In a subdivided first-stage unit, the group with the largest population share was selected with certainty and one of the remaining groups was selected by simple random sampling.  Listing and household selection were conducted independently in the two selected groups.  An undivided first-stage unit was treated as group 1.

\subsection{Second-stage strata and household selection}

Schedule 25.2 used three second-stage strata.  The household allocation retained from the project report is shown in \cref{tab:sss-allocation}.

\begin{table}[h]
\centering
\caption{Composition of second-stage strata and number of sampled households.}
\label{tab:sss-allocation}
\begin{tabularx}{0.95\textwidth}{@{}p{0.10\textwidth}Xcc@{}}
\toprule
Second-stage stratum & Household composition & No subdivision & Per selected group after subdivision \\
\midrule
SSS 1 & At least one student receiving technical or professional education & 2 & 1 \\
SSS 2 & Among the remainder, at least one student receiving general education & 4 & 2 \\
SSS 3 & Other households & 2 & 1 \\
\bottomrule
\end{tabularx}
\end{table}

Households were selected by simple random sampling without replacement within each second-stage stratum.  The allocation deliberately oversampled households with students, which is why the $H/h$ household-expansion factor must be applied separately within second-stage strata.

\section{Computational audit of the supplied R code}\label{app:audit}

The separate R file is a cleaned implementation rather than a verbatim copy of the 706-line listing in the source report.  The following changes are material for reproducibility.

The original script defined functions named \texttt{estimate\_language} and \texttt{variance\_language} more than once, so later definitions silently replaced earlier matrix-valued functions.  It also called an undefined function \texttt{estimate()}, duplicated several data-splitting blocks, omitted an argument in one urban-Himachal calculation, and contained a malformed expression in the final urban-language block.  The revised file uses unique function names and vectorized grouped calculations.

The original equality indicator converted the two labelled language variables to factors separately and then compared their numeric factor indices.  Those indices can differ when the observed level sets or their order differ, even though the displayed language labels agree.  The revised file compares normalized language labels directly.

The original script hard-coded \texttt{state\_code == "22"}.  The revised file selects Himachal Pradesh from the state label whenever possible and uses \texttt{02} only as a configurable fallback.  It stops with an informative error if the state cannot be identified and instructs the analyst to verify any recoded file.

The original variance code did not make the combined-weight versus subsample-weight convention explicit.  The revised file distinguishes the two cases, forms full-total replicate estimates accordingly, and computes the linearized ratio variance in \cref{eq:ratio-var}.  It also records active cells, paired cells, support rate, and concentration, which are required for the stability analysis.

Finally, the original listing did not contain the code that produced \texttt{Rplot.pdf}, \texttt{Rplot02.pdf}, \texttt{Rplot03.pdf}, and \texttt{Rplot04.pdf}.  Inspection of the supplied graphics identifies the correct mapping as India overall, Himachal Pradesh overall, rural Himachal Pradesh, and urban Himachal Pradesh, respectively; \texttt{Rplot01.pdf} is a duplicate India panel.  The revised script reproduces the four sample-frequency files under this mapping and also writes separate design-weighted percentage plots with confidence intervals when the private-institution indicator and reason variable are identified in the microdata.

\bibliographystyle{plainnat}
\bibliography{survey-refs}

\clearpage
\appendix
\section{External Data Tables}
\label{sec:appendix_data_tables}

\setcounter{table}{6}
\newcommand{\renderhugecsv}[3]{%
  \begingroup
  \catcode`\_=12
  \catcode`\%=12
  \catcode`\$=12
  \catcode`\&=12
  \catcode`\#=12
  \catcode`\~=12
  \catcode`\^=12
  \scriptsize 
  \setlength{\tabcolsep}{3pt} 
  \captionsetup{hypcap=false} 
  \captionof{table}{#2}\label{#3}
  \vspace{-1em}
  \addtocounter{table}{-1}
  \csvautolongtable{#1} 
  \endgroup
}

The tables in this appendix contain the estimated percentages with estimated standard deviations and coefficients of variation, generated directly from the supplementary CSV data files.

\subsection{Himachal Pradesh Demographics}

\renderhugecsv{Himachal-Pradesh-All-Female.csv}{Himachal Pradesh, female}{tab:csv_hp_all_female}
\renderhugecsv{Himachal-Pradesh-All-Male.csv}{Himachal Pradesh, male}{tab:csv_hp_all_male}
\renderhugecsv{Himachal-Pradesh-All-Neutral.csv}{Himachal Pradesh, all persons}{tab:csv_hp_all_neutral}
\renderhugecsv{Himachal-Pradesh-Rural-Female.csv}{Himachal Pradesh, rural female}{tab:csv_hp_rural_female}
\renderhugecsv{Himachal-Pradesh-Rural-Male.csv}{Himachal Pradesh, rural male}{tab:csv_hp_rural_male}
\renderhugecsv{Himachal-Pradesh-Rural-Neutral.csv}{Himachal Pradesh, rural, all persons}{tab:csv_hp_rural_neutral}
\renderhugecsv{Himachal-Pradesh-Urban-Female.csv}{Himachal Pradesh, urban female}{tab:csv_hp_urban_female}
\renderhugecsv{Himachal-Pradesh-Urban-Male.csv}{Himachal Pradesh, urban male}{tab:csv_hp_urban_male}
\renderhugecsv{Himachal-Pradesh-Urban-Neutral.csv}{Himachal Pradesh, urban, all persons}{tab:csv_hp_urban_neutral}

\clearpage
\subsection{India Demographics}

\renderhugecsv{India-All-Female.csv}{India, female}{tab:csv_ind_all_female}
\renderhugecsv{India-All-Male.csv}{India, male}{tab:csv_ind_all_male}
\renderhugecsv{India-All-Neutral.csv}{India, all persons}{tab:csv_ind_all_neutral}
\renderhugecsv{India-Rural-Female.csv}{India, rural female}{tab:csv_ind_rural_female}
\renderhugecsv{India-Rural-Male.csv}{India, rural male}{tab:csv_ind_rural_male}
\renderhugecsv{India-Rural-Neutral.csv}{India, rural, all persons}{tab:csv_ind_rural_neutral}
\renderhugecsv{India-Urban-Female.csv}{India, urban female}{tab:csv_ind_urban_female}
\renderhugecsv{India-Urban-Male.csv}{India, urban male}{tab:csv_ind_urban_male}
\renderhugecsv{India-Urban-Neutral.csv}{India, urban, all persons}{tab:csv_ind_urban_neutral}

\end{document}